\documentclass[pdflatex,sn-nature]{sn-jnl}

\usepackage{graphicx}%
\usepackage{multirow}%
\usepackage{amsmath,amssymb,amsfonts}%
\usepackage{amsthm}%
\usepackage{mathrsfs}%
\usepackage[title]{appendix}%
\usepackage{xcolor}%
\usepackage{textcomp}%
\usepackage{manyfoot}%
\usepackage{booktabs}%
\usepackage{algorithm}%
\usepackage{algorithmicx}%
\usepackage{algpseudocode}%
\usepackage{listings}%
\usepackage{lineno}

\theoremstyle{thmstyleone}%
\theoremstyle{thmstyletwo}%

\theoremstyle{thmstylethree}%

\begin{document}

\title[Article Title]{A second-generation star in a relic dwarf galaxy}


\author*[1,2]{\fnm{Anirudh}~\sur{Chiti}}\email{achiti@uchicago.edu}
\author[3]{\fnm{Vinicius}~M.~\sur{Placco}}
\author[4]{\fnm{Andrew}~B.~\sur{Pace}}
\author[1,2,5]{\fnm{Alexander}~P.~\sur{Ji}}
\author[6]{\fnm{Deepthi}~S.~\sur{Prabhu}}
\author[7]{\fnm{William}~\sur{Cerny}} 
\author[1,2]{\fnm{Guilherme}~\sur{Limberg}} 
\author[8]{\fnm{Guy}~S.~\sur{Stringfellow}}
\author[9,1,2,5]{\fnm{Alex}~\sur{Drlica-Wagner}}
\author[4]{\fnm{Kaia}~R.~\sur{Atzberger}}
\author[3]{\fnm{Yumi}~\sur{Choi}}
\author[10]{\fnm{Denija}~\sur{Crnojevi\'c}}
\author[11]{\fnm{Peter}~S.~\sur{Ferguson}}
\author[4]{\fnm{Nitya}~\sur{Kallivayalil}}
\author[12]{\fnm{Noelia}~E.~D.~\sur{No\"el}}
\author[13,14]{\fnm{Alexander}~H.~\sur{Riley}}
\author[6]{\fnm{David}~J.~\sur{Sand}}
\author[15]{\fnm{Joshua}~D.~\sur{Simon}}
\author[16]{\fnm{Alistair}~R.~\sur{Walker}}
\author[17]{\fnm{Clecio}~R.~\sur{Bom}}
\author[18]{\fnm{Julio}~A.~\sur{Carballo-Bello}}
\author[19,20]{\fnm{David}~J.~\sur{James}}
\author[21]{Clara~E.~Mart\'inez-V\'azquez}
\author[22,23]{\fnm{Gustavo}~E.~\sur{Medina}}
\author[16]{\fnm{Kathy}~\sur{Vivas}}

\affil*[1]{\orgdiv{Department of Astronomy \& Astrophysics}, \orgname{University of Chicago}, \orgaddress{\street{5640 S Ellis Ave}, \city{Chicago}, \postcode{60637}, \state{IL}, \country{USA}}}

\affil[2]{\orgdiv{Kavli Institute for Cosmological Physics}, \orgname{University of Chicago}, \orgaddress{\street{5640 S Ellis Ave}, \city{Chicago}, \postcode{60637}, \state{IL}, \country{USA}}}

\affil[3]{\orgname{NSF NOIRLab}, 
\orgaddress{Tucson, AZ 85719, USA}}

\affil[4]{\orgdiv{Department of Astronomy}, 
\orgname{University of Virginia}, 
\orgaddress{530 McCormick Road, Charlottesville, VA 22904, USA}}

\affil[5]{\orgname{NSF-Simons AI Institute for the Sky (SkAI)}, \orgaddress{172 E. Chestnut St., Chicago, IL 60611, USA}}

\affil[6]{\orgdiv{Department of Astronomy/Steward Observatory}, 
\orgname{University of Arizona}, 
\orgaddress{933 North Cherry Avenue, Tucson, AZ 85721, USA}}

\affil[7]{\orgdiv{Department of Astronomy}, 
\orgname{Yale University}, 
\orgaddress{New Haven, CT 06520, USA}}

\affil[8]{\orgdiv{Center for Astrophysics and Space Astronomy}, 
\orgname{University of Colorado Boulder}, 
\orgaddress{Boulder, CO 80309, USA}}

\affil[9]{\orgname{Fermi National Accelerator Laboratory}, 
\orgaddress{P.O.\ Box 500, Batavia, IL 60510, USA}}

\affil[10]{\orgdiv{Department of Physics \& Astronomy}, 
\orgname{University of Tampa}, 
\orgaddress{401 West Kennedy Boulevard, Tampa, FL 33606, USA}}

\affil[11]{\orgdiv{DiRAC Institute, Department of Astronomy}, \orgname{University of Washington}, 
\orgaddress{3910 15th Ave NE, Seattle, WA, 98195, USA}}

\affil[12]{\orgdiv{Department of Physics}, 
\orgname{University of Surrey}, 
\orgaddress{Guildford GU2 7XH, UK}}

\affil[13]{\orgdiv{Institute for Computational Cosmology, Department of Physics}, 
\orgname{Durham University}, 
\orgaddress{South Road, Durham DH1 3LE, UK}}

\affil[14]{\orgdiv{Lund Observatory, Division of Astrophysics, Department of Physics}, 
\orgname{Lund University}, 
\orgaddress{SE-221 00 Lund, Sweden}}

\affil[15]{\orgname{Observatories of the Carnegie Institution for Science}, 
\orgaddress{813 Santa Barbara St., Pasadena, CA 91101, USA}}

\affil[16]{\orgname{Cerro Tololo Inter-American Observatory/NSF NOIRLab}, 
\orgaddress{Casilla 603, La Serena, Chile}}

\affil[17]{\orgname{Centro Brasileiro de Pesquisas F\'isicas}, 
\orgaddress{Rua Dr. Xavier Sigaud 150, 22290-180 Rio de Janeiro, RJ, Brazil}}

\affil[18]{\orgdiv{Instituto de Alta Investigaci\'on}, 
\orgname{Universidad de Tarapac\'a}, 
\orgaddress{Casilla 7D, Arica, Chile}}

\affil[19]{\orgname{ASTRAVEO LLC}, 
\orgaddress{PO Box 1668, Gloucester, MA 01931, USA}}

\affil[20]{\orgname{Applied Materials Inc.}, 
\orgaddress{35 Dory Road, Gloucester, MA 01930, USA}}

\affil[21]{\orgname{NSF NOIRLab}, 
\orgaddress{670 N. A'ohoku Place, Hilo, Hawai'i, 96720, USA}}

\affil[22]{\orgdiv{David A. Dunlap Department of Astronomy \& Astrophysics}, 
\orgname{University of Toronto}, 
\orgaddress{50 St George Street, Toronto ON M5S 3H4, Canada}}

\affil[23]{\orgdiv{Department of Astronomy and Astrophysics}, 
\orgname{University of Toronto}, 
\orgaddress{50 St. George Street, Toronto ON, M5S 3H4, Canada}}

\abstract{
\textbf{
Stars that contain only trace amounts of elements heavier than helium, referred to as having low ``metallicity", preserve the chemical fingerprints of the first generation of stars and supernovae\cite{cds+04, bc+05, fn+15}. 
In the Milky Way, the lowest metallicity stars show an extreme over-abundance of carbon relative to other elements\cite{cgk+04, kbf+14}, which has been hypothesized to be a unique result of the first low-energy supernovae\cite{iut+05, cm+14, jfb+15}. 
However, the origin of this signature has remained a mystery, since no such stars have been discovered in the ancient dwarf galaxies where they are thought to have formed\cite{sst+15, jbb+17}.
Here, we present observations of a star in the $>10$\,billion year old ultra-faint dwarf galaxy Pictor II\cite{pace+25}, that shows the lowest iron and calcium abundances outside the Milky Way ($<$1/43,000th solar and $\sim$1/160,000th solar), with a factor of $>$3000x relative carbon enhancement.
As the first unambiguous second-generation star in a relic dwarf galaxy, this object demonstrates that carbon-enhanced second-generation stars can originate in primordial small-scale systems.
This star supports the hypothesis that carbon-enhancement is produced by low-energy-supernovae, since the yields of energetic supernovae are harder to retain in small-scale environments\cite{cm+14}.
This key local signature of chemical enrichment by the first stars traces a regime inaccessible to current high-redshift observations\cite{bsc+23}, which cannot detect the early enrichment of the smallest galaxies.
}}

\keywords{First Stars, First Galaxies, Chemical Evolution}

\maketitle

We searched for primordial, low metallicity stars in nearby ultra-faint dwarf galaxies (UFDs; stellar masses of $< 10^5$ solar mass) using imaging data from the Mapping the Ancient Galaxy in CaHK (MAGIC) Survey. 
The MAGIC Survey has imaged $\sim2000$\,deg.$^2$ of the southern hemisphere with the Dark Energy Camera/NSF Víctor M. Blanco 4\,m Telescope\cite{fdh+15}, using a narrow-band filter that covers the Ca~II~H~\&~K absorption features at $\sim$393.5\,nm.
The filter enables the identification of the lowest metallicity stars due to the strong dependence of the strength of these lines on metallicity\cite{bps+85, ksb+07, smy+17}. 

\begin{figure}[h]
\centering
\includegraphics[width=1.0\textwidth]{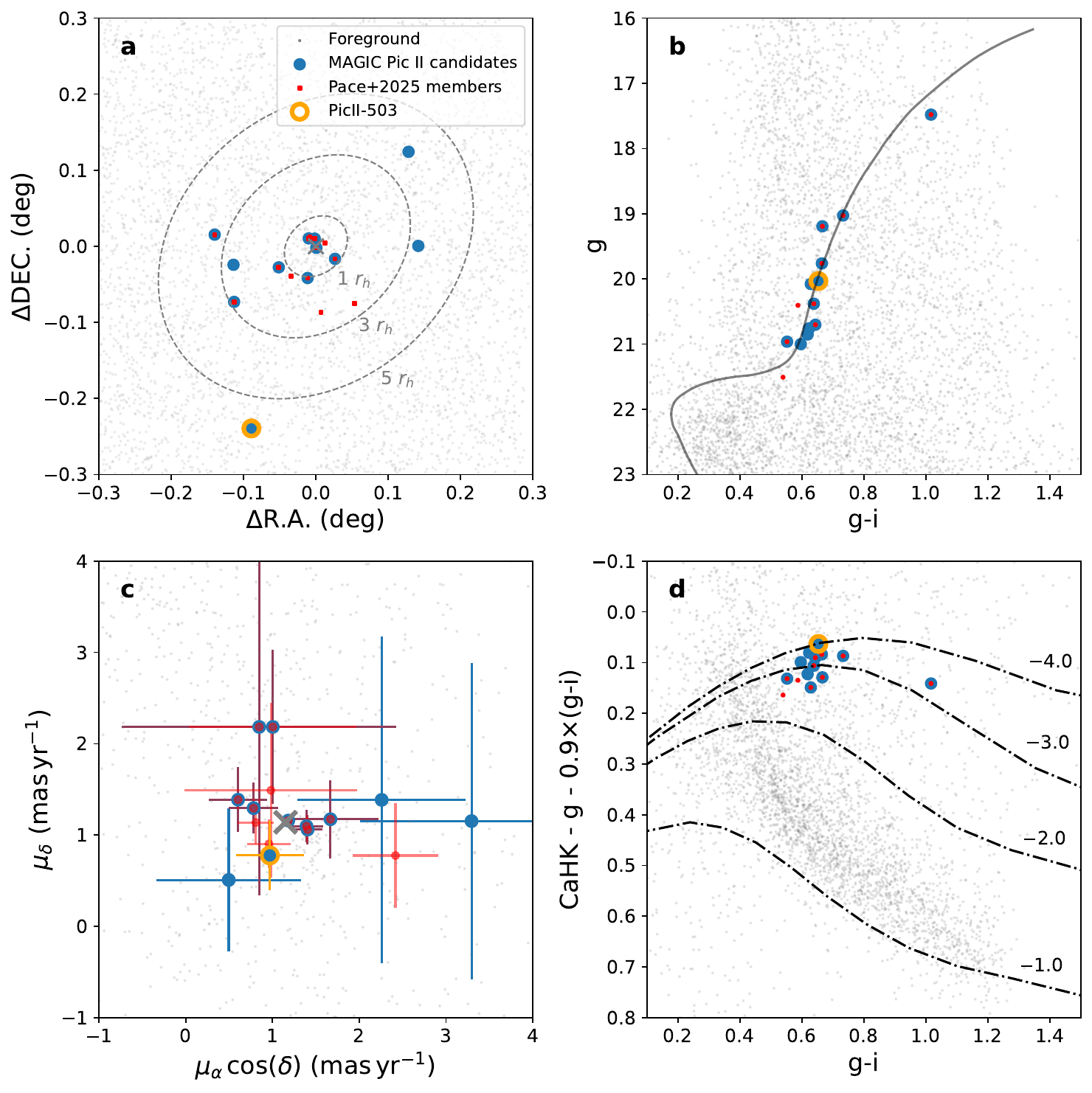}
\caption{Identification of the distant Pictor II member star, PicII-503.\\
\textbf{a.} Spatial distribution of Pictor II candidate member stars identified in this study in blue with the observed star PicII-503 circled in orange, confirmed members from ref\cite{pace+25} in red, and foreground stars in grey. 
Dashed ellipses correspond to 1, 3, and 5 times the half-light radius of Pictor II\cite{pace+25}.\\
\textbf{b.} A color-magnitude diagram of the stars, with a track (``isochrone") corresponding to a 12\,Gyr, [Fe/H] = $-2.5$ stellar population\cite{dcj+08} overlaid (solid black line) at the distance of Pictor II (45.7 kiloparsecs)\cite{pace+25}.
Some members from ref\cite{pace+25} are missing due to lacking $i$ band photometry.\\
\textbf{c.} A plot of the proper motions of the stars, with the systemic motion of Pictor II marked as a grey cross\cite{btt+22}. \\
\textbf{d.} 
CaHK$-$g$-0.9\times$(g$-$i) vs. g$-$i colour–colour diagram, used to estimate photometric metallicities. 
Contours correspond to model predictions at surface gravity logg = 2.0 (see Methods). Candidate members are selected with [Fe/H] $< -2.5$, separating them from the more metal-rich Milky Way foreground. PicII-503 (orange) lies at the lowest-metallicity end of the distribution.
}
\label{fig:selection}
\end{figure}

The Pictor~II UFD was imaged by the MAGIC Survey in February 2024. 
Figure~\ref{fig:selection} outlines our selection of its member stars (Panel a) via a color-magnitude diagram (Panel b), together with coherent proper motions from the third data release (DR3) of the \textit{Gaia} space mission (Panel c)\cite{gaia1,gaia2,btt+22}. 
Figure~\ref{fig:selection}d shows a proxy for Ca II HK strength versus broad-band color, with tracks of constant metallicity over-plotted (see methods). 
All stars with an estimated metallicity of [Fe/H]$_{\texttt{magic}}$\footnote{The notation [X/Y] is defined as the logarithmic elemental ratio of X-to-Y relative to the ratio for the same elements observed in the Sun} $< -2.5$ that passed the aforementioned criteria were considered low metallicity candidate members. 
We flagged PicII-503, the lowest metallicity candidate, for spectroscopic follow-up (Table~\ref{tab:obs}) to further investigate its nature.

\begin{figure}[h]
\centering
\includegraphics[width=1.0\textwidth]{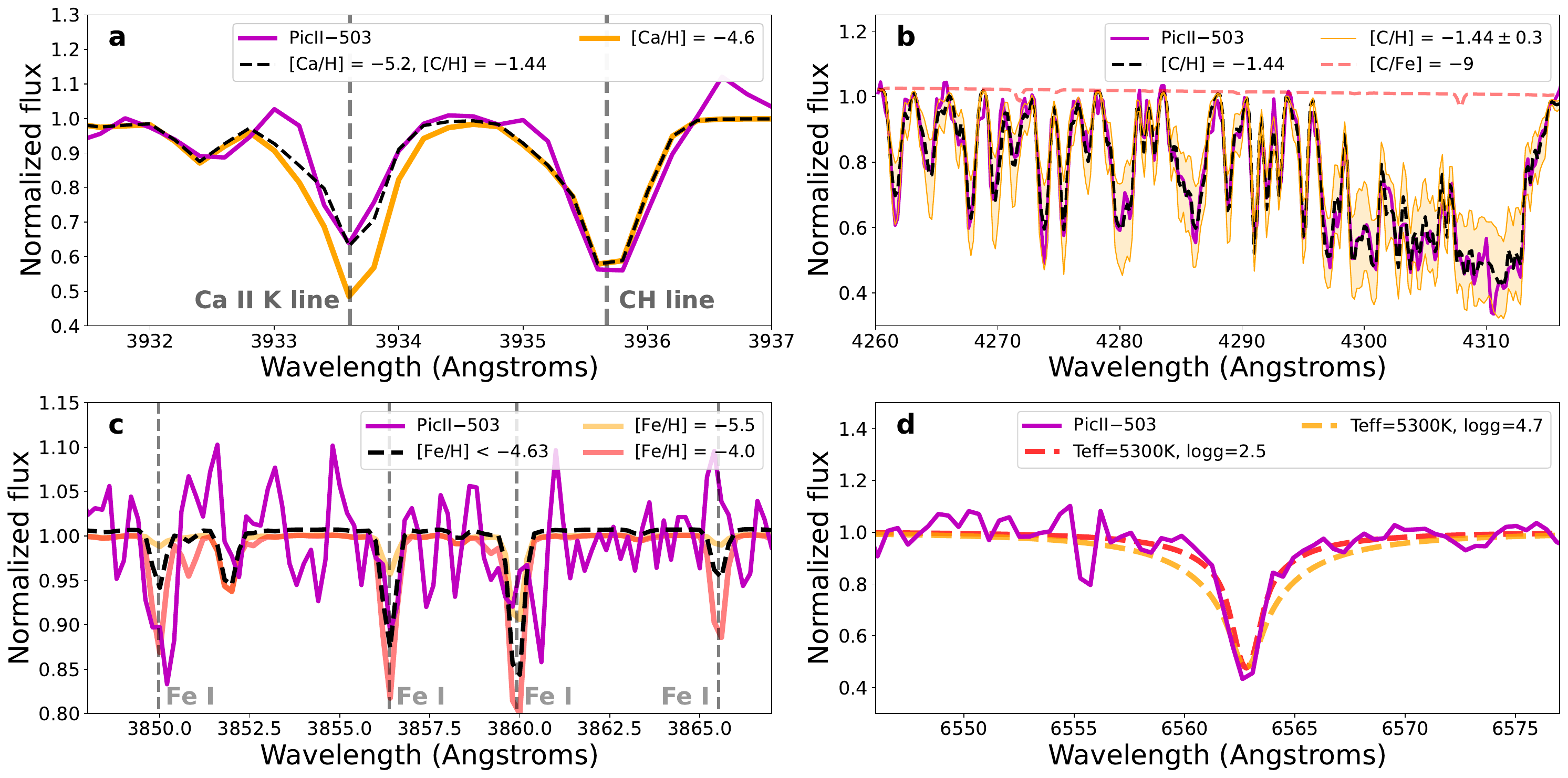}
\caption{Spectroscopic confirmation of PicII-503 as an ultra metal-poor, carbon-enhanced star.\\
\textbf{a.} A plot of the wavelength region around the calcium II K (Ca II K) absorption feature, where the observed X-Shooter spectrum is in magenta, and synthetic spectra at [Ca/H] = $-5.2$ and [Ca/H] = $-4.6$ ($\sim2\sigma$ statistical upper limit) are overplotted. 
The Ca II K feature is typically the strongest metal line in stars, and is barely detected for PicII-503.
The adjacent absorption feature is an otherwise weak carbon line that appears due to the high carbon abundance of the star.\\
\textbf{b.} The region around the molecular CH G band region in the X-Shooter data is plotted. 
An extremely prominent molecular carbon feature is detected, demonstrating the high abundance of carbon relative to other metals in the star.
A synthetic spectrum with negligible carbon abundance (red line) is shown for reference, as is a carbon synthesis at $\pm0.3$\,dex (orange band).\\
\textbf{c.} The region around the strongest Fe line at 3859\,{\AA} in the X-Shooter data is shown. A joint analysis of this feature and other weak features nearby provides a statistical upper limit of [Fe/H] $< -4.63$. 
Synthetic spectra at [Fe/H] = $-4.0$ and [Fe/H] = $-5.5$ are shown for illustrative purposes.\\
\textbf{d.} The region around the H$\alpha$ line in the MagE data is shown, with template spectra for a dwarf star ($\log\,g$ = 4.7) and a red giant star ($\log\,g$ = 2.5) overplotted. 
The observed H$\alpha$ line morphologically agrees with the giant template, confirming that PicII-503 has a surface gravity that is consistent with membership in Pictor II (see methods).
}
\label{fig:spectra}
\end{figure}

We confirmed PicII-503 as a second-generation UFD star due to its unprecedentedly low metallicity using the MagE Spectrograph\cite{mbt+08} on the 6.5\,m Magellan/Baade Telescope. 
Our MagE spectrum showed that the star is a member of Pictor~II due to its consistent radial velocity\cite{pace+25}, and we measured an extreme calcium deficiency ([Ca/H] = $-4.81^{+0.34}_{-0.49}$) and a prominent carbon enhancement ([C/Ca] = $+3.33^{+0.32}_{-0.47}$). 
We subsequently obtained deeper, higher-resolution observations with the X-Shooter spectrograph on the 8.2\,m Very Large Telescope\cite{vdd+11}, from which we derived a direct upper limit of [Fe/H] $< -4.63$, with [Ca/H] = $-5.2^{+0.31}_{-0.56}$, and [C/Ca] $= +3.76^{+0.62}_{-0.41}$ (see Figure~\ref{fig:spectra}).
These extremely low calcium and iron abundances demonstrate that the star was enriched by a single supernova from a first-generation, metal-free star\cite{cds+04, hw+10}.

As the first known second-generation star in a UFD galaxy, observations of PicII-503 inform models of the enrichment of primordial small-scale systems. 
UFDs are relic galaxies, as evidenced by their ancient stellar populations ($\gtrsim12$\, billion years old)\cite{brown+14} and very low metallicities\cite{sg+07}.
Since galaxies grow via the merging of smaller systems\cite{sz+78} and UFDs are the smallest confirmed galaxies\cite{s+19} (e.g., Pictor~II has a stellar mass of just $\sim2\times10^3$ solar masses\cite{pace+25}), it follows that UFDs are analogs of early building blocks of larger galaxies.
Accordingly, UFDs ought to trace the initial enrichment of the universe by hosting stars enriched by the first supernovae.
PicII-503, with an unprecedentedly low metallicity of [Fe/H] $< -4.63$, is the first observational confirmation of this initial enrichment; all other known UFD stars have [Fe/H]~$>~-4.0$\cite{bcf+25}.  

\begin{figure}[h]
\centering
\includegraphics[width=0.9\textwidth]{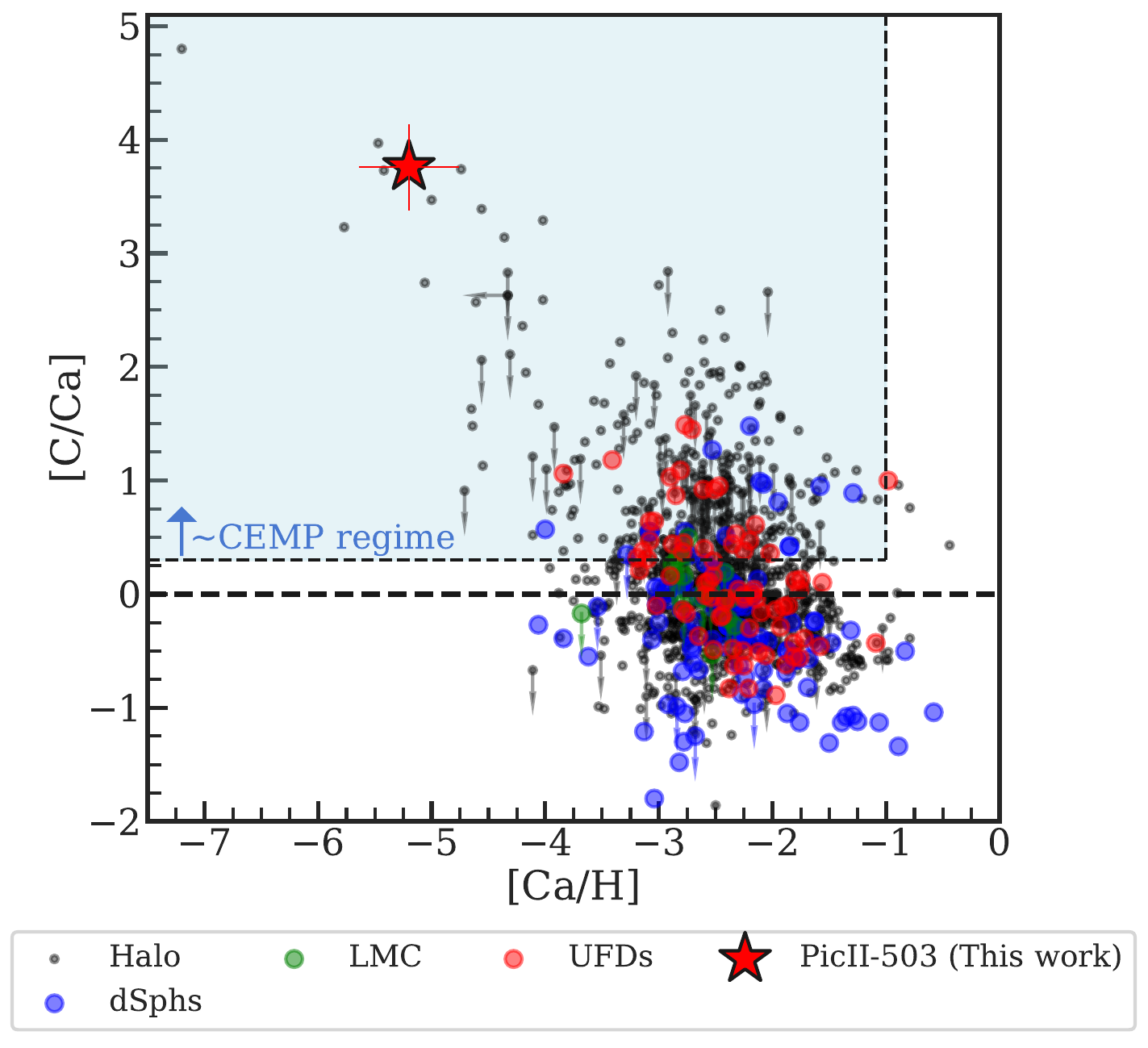}
\caption{A plot of the carbon-to-calcium ratio versus calcium abundance, including our observed star in Pictor II (red star), along with Milky Way halo stars (black circles), other stars in UFDs (red circles), and stars in larger dwarf galaxies (dSphs; blue) and the LMC (green).
Carbon-enhancement in the Milky Way is defined to occur when [C/Fe] $> 0.7$.
Given that calcium behaves as an $\alpha$-element, and [$\alpha$/Fe] $\approx$ 0.4 at low metallicities due to dominant core collapse supernova enrichment\cite{ww+95}, we indicate the regime where [C/Ca] $> 0.3$ as a proxy for the carbon-enhanced regime.
The dramatic carbon-enhancement in the observed Pictor II star is evident, placing it among the most carbon-enhanced and metal-deficient stars known, demonstrating that these second generation stars could have originated in relic dwarf galaxies. 
}
\label{fig:abundances}
\end{figure}

Notably, PicII-503's element abundances match the canonical signature of stars with the lowest iron abundances in the Milky Way halo-- a dramatic enhancement in carbon (see Figure~\ref{fig:abundances})\cite{bc+05}.
These stars are known as carbon-enhanced metal-poor (CEMP) stars and it has long been suggested that they preserve a key nucleosynthetic signature of the first stars\cite{cm+14}.
The origin of CEMP stars has remained a mystery, with competing interpretations\cite{iut+05,mem+06}.
One hypothesis is that these stars preserve enrichment by low-energy supernovae, in which heavy elements that form close to the interior of the star (e.g., Fe) fall back into the remnant compact object, while lighter elements from the star’s outer regions (e.g., C, N, O) are ejected and enrich the early interstellar medium\cite{iut+05}.
Other explanations point to winds from massive stars that produce large C, N, O enhancements\cite{mem+06} or cooling threshold arguments\cite{bl+03, fjb+07, hy+19}.

PicII-503 demonstrates that second-generation CEMP stars in the Milky Way halo can originate from accreted relic galaxies, and supports a low-energy-supernova origin of their enrichment.
The low-energy-supernova scenario predicts that second-generation CEMP stars preferentially form in low-mass galaxies.
This dependence occurs because supernovae that produce normal carbon signatures (e.g., hypernovae\cite{hw+10,ssa+21,svs+24}) have explosions that would escape from the weak gravitational potential wells of small galaxies\cite{cm+14}. 
The other theories of CEMP star formation do not have an obvious environmental dependence.
The handful of lowest metallicity stars in larger dwarf galaxies and the Milky Way bulge do not show as extreme carbon enhancement\cite{hck+15,jnm+15,asa+21,svs+24,cml+24,sav+24}, yet the fraction of stars that are carbon enhanced in UFDs appears to match or be higher than the Milky Way halo\cite{jls+20,ljs+24}.
PicII-503, as an unambiguous second-generation CEMP star in a UFD, functions as a rosetta stone that shows this peculiar first-star enrichment signature observed in the Milky Way halo can originate from primordial small-scale galaxies. 

The non-detection of barium ([Ba/H] $< -3.84$) and magnesium ([Mg/H] $< -4.42$) place further constraints on the formation of PicII-503. 
Another origin scenario for extreme carbon enhancement is the accretion of material from a binary companion star on the asymptotic giant branch (AGB)\cite{ran+05}.
However, this would produce a substantial abundance of elements from the slow neutron-capture process (s-process), for which barium is a tracer element.
The barium upper limit for PicII-503 is below what is seen for nearly all s-process enhanced CEMP stars in the Milky Way halo (see methods)\cite{af+18}.
A handful of CEMP stars show extreme magnesium enhancements (e.g., [Mg/Fe] $\gtrsim 1.5$\cite{anr+02, kbf+14}) that map to higher supernovae energies\cite{iut+05,hw+10}.
PicII-503 does not show such an enhancement in its magnesium abundance.
Comparing the measurable element abundances in PicII-503 (Mg, Fe, Ca, Ba, C, Sr, N) to supernova yield models\cite{hw+10} places an upper limit on the explosion energy of its progenitor supernova of $< 2\times$10$^{51}$ \,erg and a mass limit on the progenitor star of $<45$\,M$_\odot$. 
If a fiducial [Ca/Fe] = 0.4\cite{ww+95,fn+15} is adopted to constrain the iron abundance, the resulting fit strongly prefers a progenitor supernova energy of $3\times10^{50}$\,erg and a progenitor star mass of 12\,M$_\odot$.
Either outcome excludes enrichment by high-energy supernovae that require galaxies to have dark matter masses $\gtrsim6.3\times10^6$\,M$_\odot$ at the time of first-star formation in order for chemical yields to be retained\cite{cm+14}. 
The outcome when assuming [Ca/Fe] = 0.4 corresponds to a faint supernova that can have yields retained by even the smallest galaxies capable of forming second-generation stars (dark matter mass $\sim10^6$M$_\odot$).

Constraining the origin of the lowest metallicity, extremely carbon-enhanced stars in the Milky Way establishes a key bridge between local and high-redshift studies of the first stars. 
Current high-redshift observations, even with the James Webb Space Telescope (JWST), cannot probe the chemical signatures of gas in UFD-mass galaxies at their initial enrichment\cite{nmk+24}. 
The canonical local signature of initial enrichment by the first stars (i.e., carbon-enhancement) thus probes a different qualitative regime (e.g., low energy, faint supernovae) than what is seen in JWST observations.
The latter probes more massive galaxies at high redshifts, some of which have nitrogen enhancements\cite{bsc+23, ckr+23} that may also point to a different  enrichment process. 

Finally, we note that PicII-503 was discovered in the outskirts of this UFD (i.e., at $\gtrsim5$ times its half-light radius).
Over 450 stars have metallicities derived in UFDs (see methods); however, the vast majority of them are located in the central regions of these systems, where it is easiest to target members. 
Initial studies have suggested that lower metallicity stars may be preferentially located in the outskirts of UFDs\cite{cfs+21,lja+22}.
Our discovery of PicII-503 further supports the indication that signatures of early chemical enrichment are located in the peripheries of the faintest galaxies. 
Explaining this behavior is an intriguing puzzle for theories of early galaxy formation, with potential implications for the environments of the first stars\cite{bem+25}. 
The rarity of stars like PicII-503 has historically inhibited their identification, but we have shown that additional second-generation stars potentially await discovery in UFDs through precision photometric targeting. 
This new discovery space is thus open for upcoming facilities and thirty meter-class telescopes to probe how the first stars formed and influenced the faintest galaxies.

\begin{table}[h]
\caption{Observations \& Properties of PicII-503}\label{tab:obs}
\begin{tabular}{@{}lllllll@{}}
\toprule
Name & R.A. & Decl. & Observation Date & $g_{\text{DES}}$ & HRV$^\dagger$ &\\
& (hh:mm:ss) & (dd:mm:ss) & & (mag) & (km\,s$^{-1}$)\\
\midrule
PicII-503    & 06:43:59.01   & $-60$:08:25  & 02 Jan 2025 (MagE); & 20.35 & $331.7 \pm 3.1$ \\
 &  &  & 21-30 Mar 2025 (X-Shooter) &  & $328.9 \pm 3.7$ \\
\botrule
\end{tabular}
$^\dagger$ Heliocentric Radial Velocity (HRV)
\footnotetext{Note: The reported heliocentric radial velocity (HRV) uncertainty corresponds to a 1\,$\sigma$ value that includes statistical and systematic uncertainties.}
\end{table}

\begin{table}[h]
\caption{\label{tab:abundances}Stellar parameters and abundances of PicII-503 from X-Shooter data}\label{fig:pic2pr503}
\begin{tabular*}{\textwidth}{@{\extracolsep\fill}lccccccccccc}
\toprule%
Name &  $T_{\text{eff}}$ & $\log\,g$ & [Fe/H] & [Ca/H] & [C/H] & [C/Ca] & [Mg/H]\\
 &  (K) & (dex) &  &  &  &  & \\
\midrule
PicII-503  & 5320$\pm$150 & 2.71$\pm$0.30 & $<-4.63$ & $-5.2^{+0.31}_{-0.56}$ & $-$1.44$\pm$0.38 & $3.76^{+0.62}_{-0.41}$ & $<-4.42$\\
\midrule
 &   &  & [Ba/H] & [Sr/H] & [N/H] &  & \\
\midrule
  & &  &  $< -3.84$ & $< -4.38$ & $<-2.08$  &  & \\
\botrule
\end{tabular*}
\footnotetext{Note: The reported chemical abundance uncertainties correspond to 1\,$\sigma$ values that include statistical and systematic uncertainties.
The quoted upper limits are statistical 99.7\% one-sided limits.}
\end{table}

\clearpage

\section*{Methods}

\subsection*{Identification of PicII-503}

PicII-503 was identified in a search for the lowest metallicity stars in UFDs using data from narrow-band Ca II HK imaging (hereafter CaHK) from the DECam MAGIC survey.
MAGIC is an ongoing program on DECam that will image $\sim$5,000\,deg.$^2$ of the southern hemisphere upon survey completion in 2026.
The survey will provide the deepest CaHK imaging of this region to date, enabling efficient identification of ultra metal-poor stars in the Milky Way halo and a number of satellite galaxies.
The survey strategy and early science will be described in an upcoming paper (Chiti et al., in prep), but initial MAGIC metallicity validation has been performed with data on the Sculptor dwarf galaxy\cite{bcl+25}, and MAGIC data has been used to target low metallicity outer halo stars\cite{plc+25} and confirm members of the Pictor II UFD\cite{pace+25}.
Briefly, metallicities from this data are derived closely following the methods in refs\cite{cfj+20, cfm+21}, in which observed values for stars in the color-color space shown in Figure~\ref{fig:selection}d are compared to synthetic photometry for stars spanning a range of stellar parameters ($-0.5 < \log\,g < 5.5$, $-5.0 <$ [Fe/H] $ < +1$, 2500\,K $< T_{\text{eff}} < $8000\,K). 
The grid of synthetic photometry was derived from a grid of flux-calibrated synthetic spectra that was generated from the Turbospectrum code\cite{ap+98,p+12}, MARCS model atmospheres\cite{gee+08}, and a linelist from the Vienna Atomic Line Database (VALD)\cite{pkr+95, rpk+15}.

Candidate members of the Pictor~II UFD were identified following a selection along the color-magnitude diagram of the galaxy, finding stars with consistent proper motions in \textit{Gaia} DR3\cite{gaia2,gaiaastrometry}, and by selecting stars with CaHK photometric metallicities [Fe/H]$_{\text{magic}}~<~-2.5$ to exclude metal-rich Milky Way foreground stars.
For the color-magnitude diagram selection, we adopted a Dartmouth isochrone of 12\,Gyr in age and [Fe/H] $= -2.5$, [$\alpha$/Fe] = $= 0.4$\cite{dcj+08}, placed at the distance of Pictor II\cite{dba+16} and selected stars along this track (Figure~\ref{fig:selection}, Panel b). 
For the proper motion selection, we selected stars with proper motions consistent within 2.5\,$\sigma$ of the systemic proper motion of Pictor II ($\mu_{\alpha}\cos\,\delta$ = 1.15\,mas\,yr$^{-1}$, $\mu_{\delta} = $1.14\,mas\,yr$^{-1}$)\cite{btt+22, pel+22}.
After applying the photometric metallicity selection, we recovered the on-sky sample of candidate members shown in panel a of Figure~\ref{fig:selection}. 
We note that two members in ref\cite{pace+25} show up as low metallicity stars in Figure~\ref{fig:selection}d but were not included in our selection. 
This is due to one star not having a proper motion in \textit{Gaia} DR3, and the other having a proper motion ($\mu_{\alpha}\cos\,\delta$ = 2.42$\pm0.50$\,mas\,yr$^{-1}$, $\mu_{\delta} = $0.78$\pm0.58$\,mas\,yr$^{-1}$) marginally more than 2.5\,$\sigma$ from the Pictor II proper motion.
PicII-503, the most distant candidate in the plot, also had the lowest photometric metallicity of the sample ([Fe/H]$_{\text{magic}} = -3.82\pm0.65$). 
Accordingly, we flagged it for high-priority spectroscopic follow-up observations.

We observed PicII-503 with the MagE spectrograph on January 2, 2025 for three exposures totaling 155\,min with the 0.7" slit in good seeing conditions ($\sim$0.7''). 
This configuration provided a resolution of $R\sim6000$ spanning a wavelength range of 310\,nm to 1000\,nm. 
The data were reduced with the CarPy reduction software\cite{k+03}.
From this data, we confirmed PicII-503 as a member using velocities from the H$\alpha$ line at 656.5\,nm and performed an initial abundance analysis to derive a calcium abundance of [Ca/H] = $-4.81^{+0.34}_{-0.49}$ from the Ca II K line at 393.4\,nm and a prominent carbon enhancement of [C/Ca] = $3.33^{+0.32}_{-0.47}$ from the molecular CH G band at $\sim430$\,nm (see subsequent methods sections).
However, the data quality only permitted a weak upper limit on the iron abundance ([Fe/H] $< -3.5$) due to a function of the resolution and low signal-to-noise (S/N$\sim9$ at $\sim380$\,nm).

We obtained additional data from the X-Shooter spectrograph consisting of 13 exposures of 45\,mins each taken between March 21–30, 2025 through Director's Discretionary Time. 
The data were obtained in reasonable seeing conditions ($< 0.9''$). We used data from the UVB arm (300\,nm to 559.5\,nm) collected with a 0.5" slit to provide a resolution of $R\sim9700$. Data were reduced with the standard EsoReflex reduction pipeline\cite{frb+13}.
The final stacked spectrum had a significantly higher signal-to-noise ratio (e.g., S/N$\sim23$ at 380\,nm) than the MagE data.
This increased S/N and higher resolution in the X-Shooter data allowed for stronger upper limits on element abundances relative to our initial MagE spectrum.

\subsection*{Radial velocity analysis}

In this section, we describe our radial velocity analysis of the MagE and X-Shooter spectroscopy of PicII-503.
For the MagE data, we derived a radial velocity by cross-correlating the reduced spectrum of PicII-503 with a MagE template spectrum of the metal-poor red giant star HD122563 that is taken to be at $-26.51$\,km\,s$^{-1}$\cite{cmf+12}.
This spectrum of HD122563 was obtained in the same configuration as our spectrum of PicII-503, and is identical to the template used in prior published velocity analysis with MagE data\cite{cfs+21,ocs+24}.
In contrast to ref\cite{cfs+21}, we performed a cross-correlation over the H$\alpha$ feature (654\,nm to 658\,nm) as opposed to the magnesium b region (490\,nm to 540\,nm), given the lack of metal lines in the spectrum of PicII-503.
We derived a velocity correction for mis-centering of PicII-503 in the MagE slit by  cross-correlating the telluric feature at 758\,nm to 771\,nm with a template spectrum of the hot star HR 4781, again following literature work\cite{sld+17,cfs+21}. 

Radial velocity uncertainties for our MagE spectrum were derived in the same manner as prior work with the DEIMOS, IMACS and MagE spectrographs\cite{sg+07, sld+17, lsd+17, cfs+21}, in which statistical uncertainties were determined by adding noise to data based on the measured signal-to-noise ratio 500 times, re-deriving the velocity, and taking the standard deviation of the resulting distribution as the uncertainty.
To derive the systematic velocity uncertainty, we repeated our routine on individual exposures of PicII-503 and the MagE data in ref\cite{cfs+21} and found that an uncertainty floor of 1.7\,km\,s$^{-1}$ needed to be added in quadrature for self-consistent uncertainties.
This process results in a radial velocity from the MagE spectrum of $331.7 \pm 3.1$\,km\,s$^{-1}$.
For the X-Shooter data, we derived the radial velocity of each individual exposure by cross-correlating the H$\beta$ feature (481\,nm to 491\,nm) with a template MIKE spectrum of HD122563 that was smoothed to the resolution of our X-Shooter spectrum ($R\sim9700$). 
The range of radial velocities across exposures was 326.3\,km\,s$^{-1}$ to 337.5\,km\,s$^{-1}$, with a mean and standard deviation of $328.9\pm3.7$\,km\,s$^{-1}$, which we adopt as the X-Shooter radial velocity measurement and uncertainty. 
Based on the X-Shooter and MagE velocities, we see no evidence for binary radial velocity variation over $sim$90 days.

\subsection*{Chemical abundance analysis}

The chemical abundances of PicII-503 were derived following standard stellar chemical abundance analysis procedures with the assumption of 1D local thermodynamic equilibrium.
Specifically, we generated synthetic stellar spectra for comparison to the observed spectra using the 2017 version of the MOOG radiative transfer code\cite{s+73,sks+11}\footnote{https://github.com/alexji/moog17scat}, ATLAS9 stellar model atmospheres\cite{ck+03,k+05}, and a linelist compiled from various sources\cite{mpv+14, slr+14, rdl+14, drl+14, bpr+17,phn+17,nist,dls+21} using the \texttt{linemake} code\cite{psr+21}\footnote{https://github.com/vmplacco/linemake}.
All abundance analysis was performed within the Spectroscopy Made Harder (SMHR) graphical wrapper\cite{c+14}\footnote{https://github.com/andycasey/smhr}.
The stellar parameters of PicII-503 (effective temperature $T_{\text{eff}}$, surface gravity $\log\,g$, and microturbulence $v_{\text{mic}}$) are required to generate synthetic spectra.
The $T_{\text{eff}}$ and $\log\,g$ were derived by matching the observed $g-i$ color of PicII-503 to the Dartmouth [Fe/H] = $-2.5$, 12\,Gyr isochrone shown in Panel b of Figure~\ref{fig:selection}.
We opt for these Dartmouth isochrones over other models due to the agreement of the former with the observed Pictor II red giant branch.
The microturbulence was adopted as 1.6\,km\,s$^{-1}$ as is seen in stars at similar $\log\,g$\cite{fcj+13}.
Fiducial uncertainties of 150\,K, 0.3\,dex, and 0.3\,km\,s$^{-1}$ were adopted for $T_{\text{eff}}$, $\log\,g$, and $v_{\text{mic}}$ respectively, following literature work\cite{fcj+13, cfs+21}.

Our two detected element abundances in PicII-503 are calcium and carbon, which were derived using the Ca II K line (393.4\,nm) and molecular carbon CH G band ($\sim430$\,nm), respectively.
For each of these features, we identified the best-fitting synthetic spectrum as the element abundance was varied following the procedure in ref\cite{jlh+20}. 
The best-fitting synthetic spectra for the Ca II K line and the CH G band are shown in Panels a and b of Figure~\ref{fig:spectra}.
Statistical uncertainties were derived by varying the abundance of the synthetic spectrum to encompass the noise in the absorption feature. Uncertainties from the stellar parameters were propagated by varying each parameter by its uncertainty, and adding the resulting shift in the abundance in quadrature to the statistical uncertainty.
In this analysis, we do not analyze the Ca II H line (396.9\,nm) as it is blended with the nearby prominent Hydrogen H$\epsilon$ line (397.1\,nm).

\begin{figure}[h]
\centering
\includegraphics[width=1.0\textwidth]{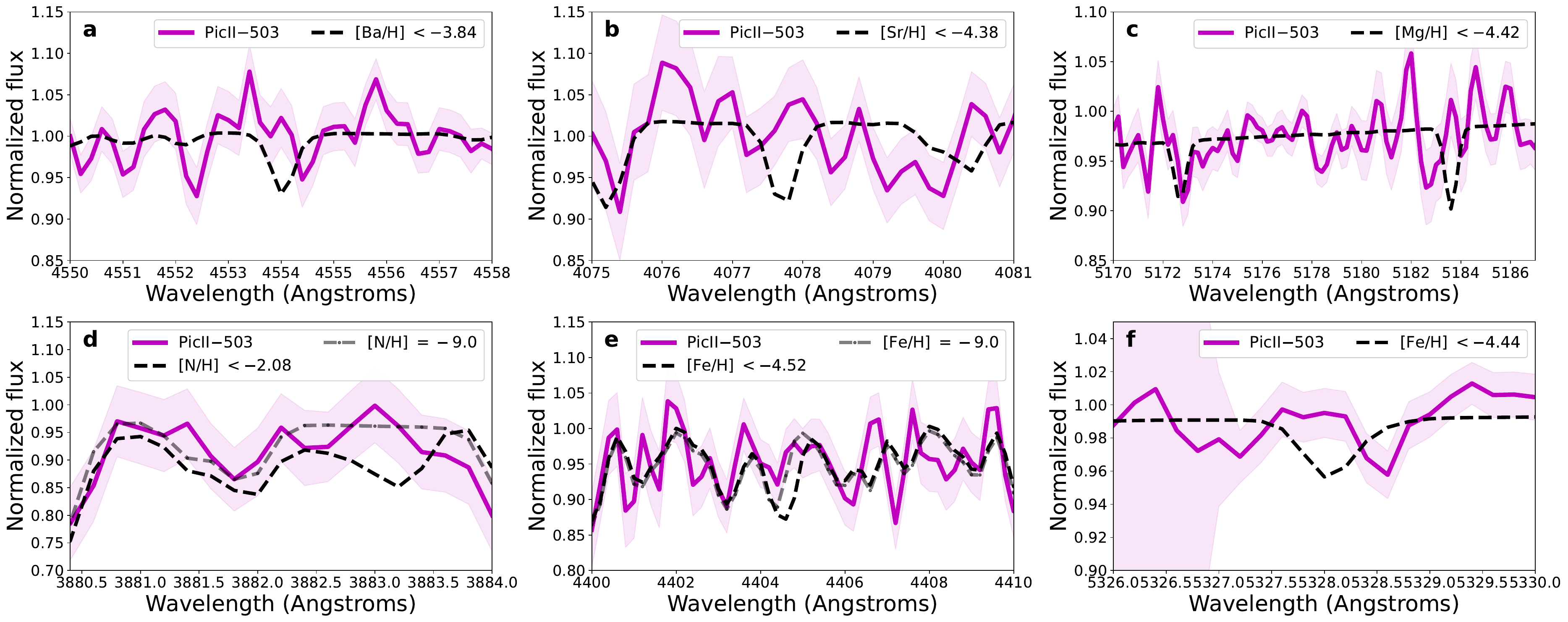}
\caption{Analysis of various absorption features in the spectrum of PicII-503.\\
\textbf{a. -- f.} Cutouts from the X-Shooter spectrum of PicII-503 around the wavelength regions of prominent element absorption features. 
Magenta shaded regions correspond to the uncertainty inferred from the S/N of the data (see methods).
Dashed lines correspond to synthetic spectra at various element abundances.
Panels a through d correspond to the abundance upper limits in Table~\ref{tab:abundances} for barium, strontium, magnesium, and nitrogen. 
In panel d, a spectrum with [N/H] = $-9.0$ is also shown to illustrate the regions that have nitrogen absorption features.
Panels e and f show regions around other prominent iron absorption features with weaker upper limits to corroborate the exceptionally low iron abundance of PicII-503.
}
\label{fig:allsynth}
\end{figure}

For all other elements (Fe, Mg, Ba, Sr, N), we derive formal 3$\sigma$ upper limits by fitting an abundance to the noise in the spectrum through a $\chi^2$ minimization, and then increasing the abundance until $\Delta\chi^2=9$ to correspond to a 3$\sigma$ increase, otherwise following the procedure in ref\cite{jls+20}.
To account for sources of systematic variations in the data that are not encompassed by the variance spectrum returned by EsoReflex, we empirically derive the S/N for each pixel by taking the standard deviation of the continuum-normalized spectrum in a rolling 20 pixel window (shown in light pink in extended data Figure~\ref{fig:allsynth}). 
This likely over-estimates the noise in the spectrum in the presence of real absorption features, but ought to be an approximate or conservative assumption when deriving abundance upper limits.
Absorption features and synthetic spectra used for the upper limits in Table~\ref{tab:abundances} are shown in Extended Data Figure~\ref{fig:allsynth}, in addition to syntheses around the location of other iron lines to corroborate our upper limit in Panel c of Figure~\ref{fig:spectra}.
Independent of our [Fe/H] upper limit from the data, the [Ca/H]$=-5.2^{+0.31}_{-0.56}$ of PicII-503 supports an iron abundance of [Fe/H] $\lesssim -4.5$.
No star in the JINAbase compilation\footnote{https://jinabase.pythonanywhere.com} of metal-poor stars has [Ca/Fe] $< -0.66$\cite{af+18}.

\subsection*{Membership confirmation}

\begin{figure}[h]
\centering
\includegraphics[width=0.49\textwidth]{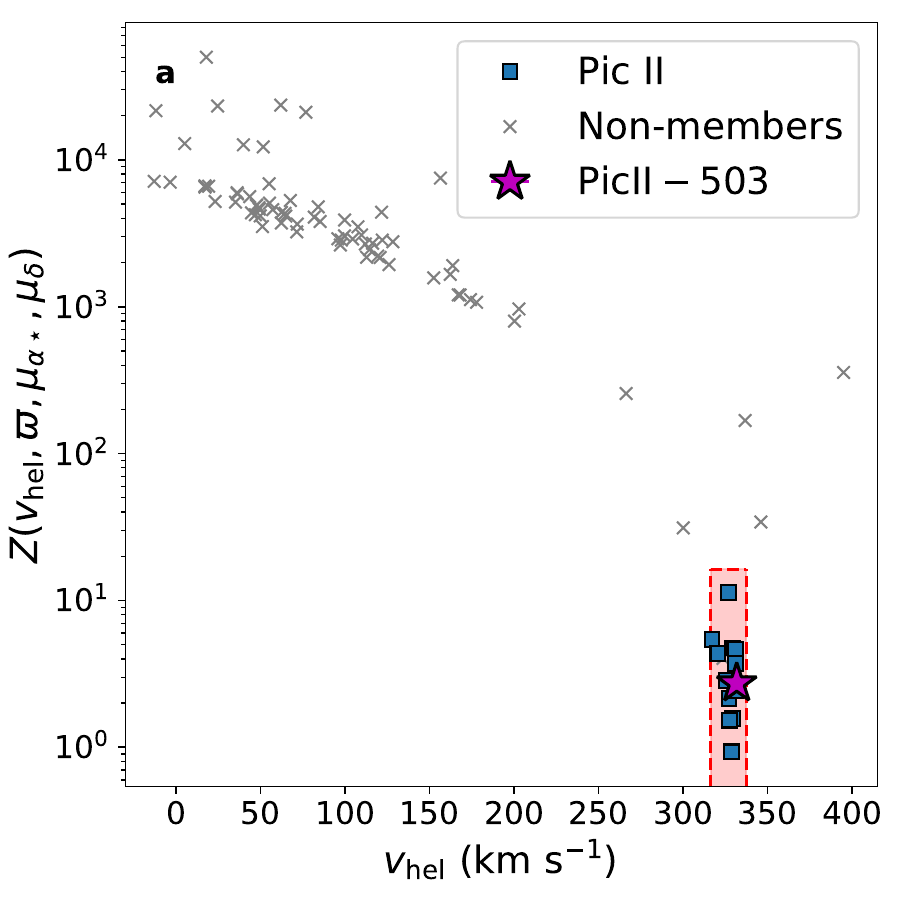}
\includegraphics[width=0.49\textwidth]{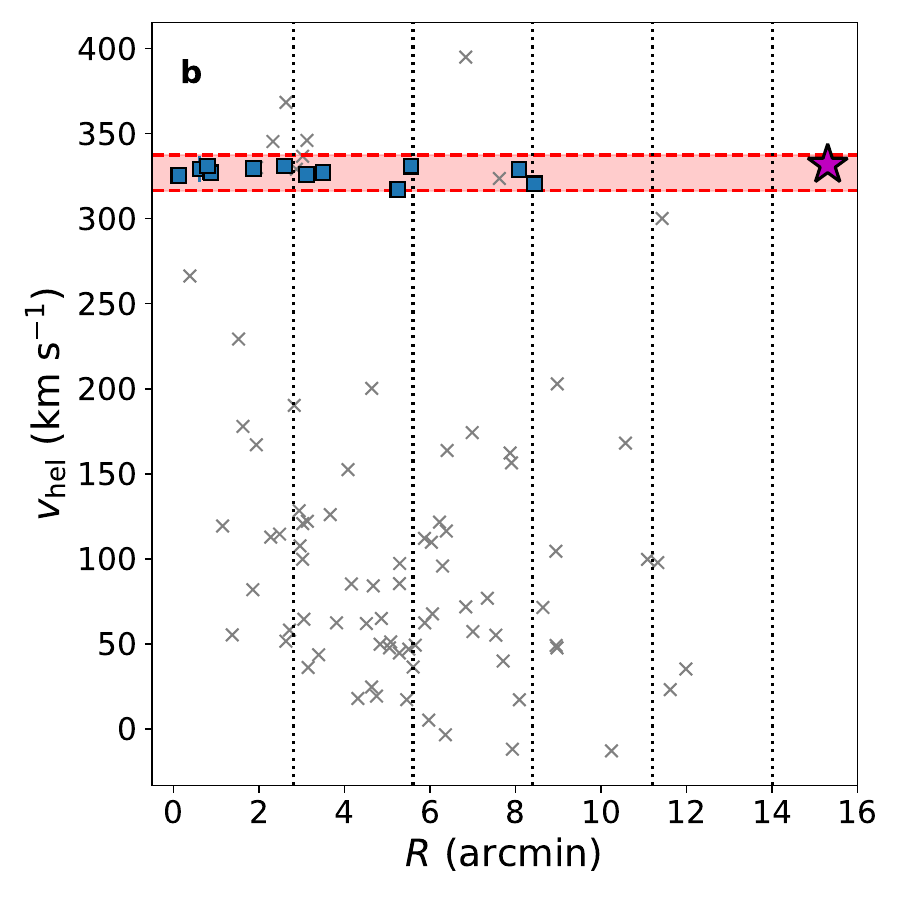}
\caption{Confirming that PicII-503 is a member of the Pictor II UFD.\\
\textbf{a.} A plot of the membership score versus radial velocity, for stars in the recent spectroscopic study of Pictor II\cite{pace+25} and PicII-503 overplotted (purple star). 
Non-members of Pictor II are shown in grey, and members are blue squares. 
Note that the systemic velocity of Pictor II is well-separated from the radial velocities of foreground stars.
The red region corresponds to $Z < 16.3$ and three times the velocity dispersion (3$\times$3.5\,km\,s$^{-1}$) around the systemic velocity (326.9\,km\,s$^{-1}$) of Pictor II\cite{pace+25}.
The former are the thresholds for velocity, proper motion, and parallax consistency with Pictor II.\\
\textbf{b.} Radial velocities of stars in the left panel versus their distance from the center of Pictor II.
One and two half-light radii are denoted by vertical dotted lines.
PicII-503 appears as the most distant likely member.
The red region corresponds to the same velocity selection as in panel a.
}
\label{fig:membership}
\end{figure}

We confirm PicII-503 as a member of the Pictor II UFD through its consistent radial velocity, proper motion, low metallicity, and classification as a red giant star. 
To first order, the systemic velocity of Pictor II is $326.9\pm1.1$\,km\,s$^{-1}$\cite{pace+25}, which is well-separated from the velocity distribution of Milky Way foreground stars in its vicinity (see Extended Data Figure~\ref{fig:membership}) and consistent with PicII-503. 
The proper motion of PicII-503 in \textit{Gaia} DR3 ($\mu_{\alpha}\cos\,\delta = 0.97\pm 0.39$\,mas\,yr$^{-1}$, $\mu_{\delta} = 0.78\pm0.38$\,mas\,yr$^{-1}$) is consistent with the systemic motion of Pictor II ($\mu_{\alpha}\cos\,\delta$ = 1.15\,mas\,yr$^{-1}$, $\mu_{\delta} = $1.14\,mas\,yr$^{-1}$)\cite{btt+22}.
Additionally, the metallicity of PicII-503 is well below what is expected for the typical metallicity distribution of stars in the Milky Way halo, which peaks from [Fe/H]$\approx$~$-1.2$~to~[Fe/H]$\approx$~$-1.8$\cite{cnz+19, ysm+21, cmf+21}.

To formalize our analysis, we compute a membership score for PicII-503 following the methodology in ref\cite{tsa+23}, jointly using \textit{Gaia} DR3 astrometry and its radial velocity.
This approach was also used in the recent spectroscopic study of Pictor II\cite{pace+25} to classify members, and the resulting analysis strongly clusters PicII-503 among the other Pictor II members (see Extended Data Figure~\ref{fig:membership}). 
As an independent check, we also plot the H$\alpha$ absorption line of PicII-503 in our MagE spectrum in Panel d of Figure~\ref{fig:spectra}, along with template spectra\cite{bpo+00, bsa+02}\footnote{https://github.com/barklem/public-data} at the approximate surface gravity of PicII-503 if it were a member ($\log\,g$ = 2.5), and the surface gravity if it were instead a main-sequence star at the same $g-i$ color ($\log\,g$ = 4.7). 
The observed spectrum is consistent with the red giant template, corroborating all other indications that the star is a member. 

\subsection*{Supernova yield fitting}

\begin{figure}[h]
\centering
\includegraphics[width=0.49\textwidth]{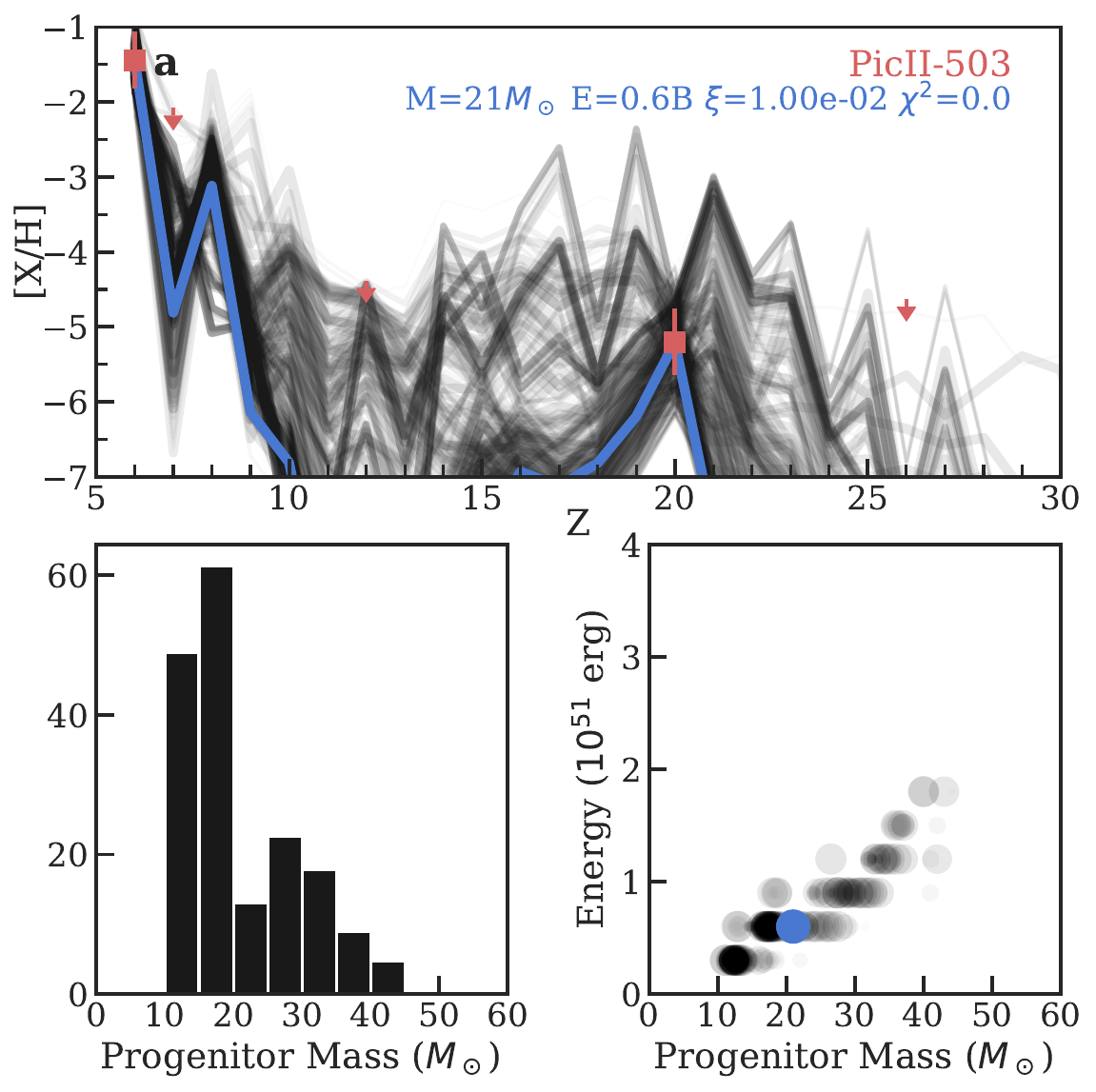}
\includegraphics[width=0.49\textwidth]{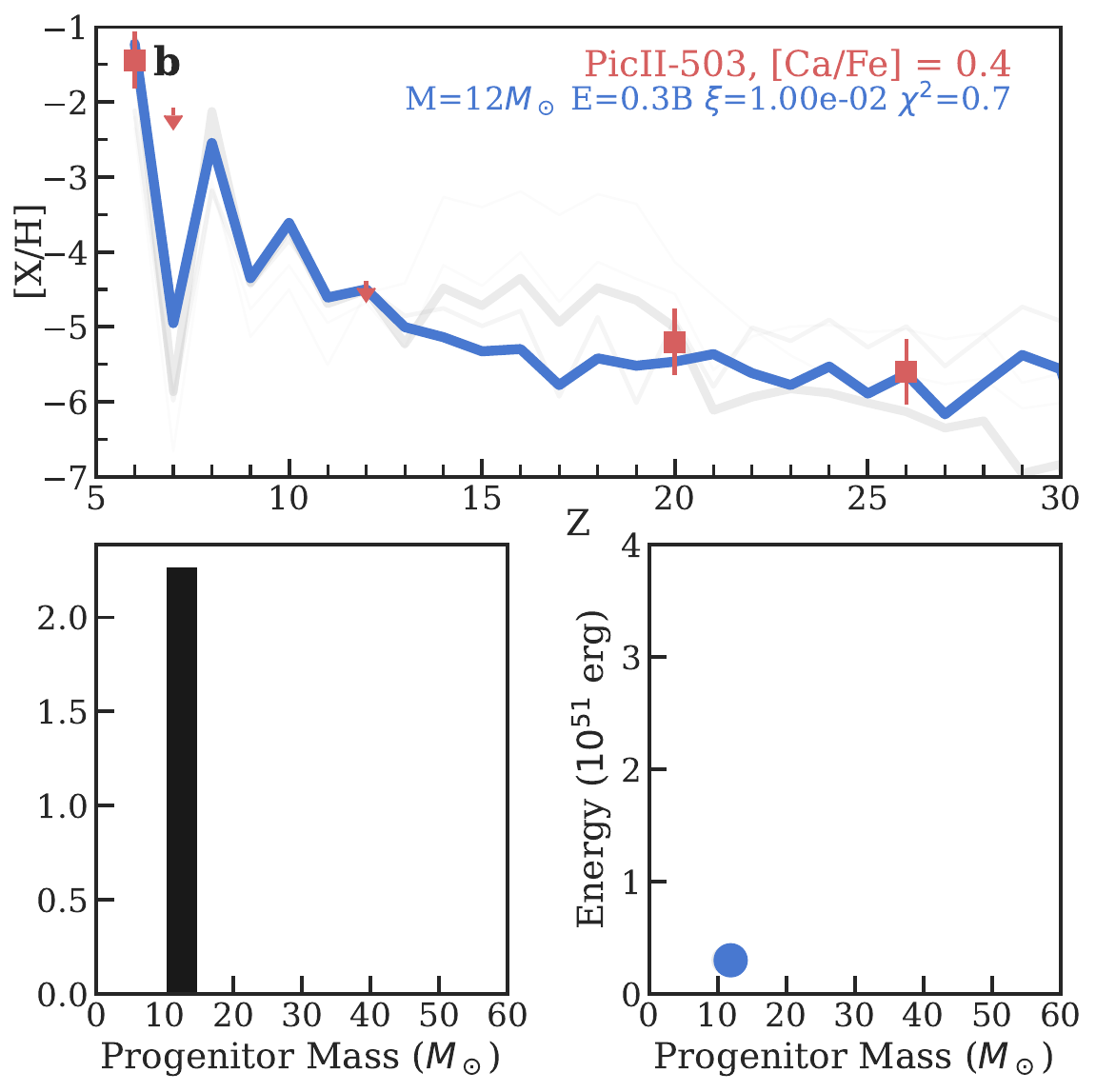}
\caption{Constraining the properties of the first-generation supernova that enriched PicII-503.\\
\textbf{a.} The top panel shows the derived element abundances of PicII-503 (red squares for detections, red arrows for upper limits), the best-fitting yield model from ref\cite{hw+10} (blue), and individual models preferred by the abundances (black lines) following the methodology described in the methods section.
For reference, calcium has atomic number Z=20, and iron has Z=26.
The model fit parameters are initial mass, energy, and a mixing parameter, which we denote with M, E, $\xi$, respectively. 
Models with worse $\chi^2$ values are plotted more transparently. 
The bottom left panel is a weighted histogram of resulting model progenitor masses, and the bottom right is a plot of model explosion energy versus progenitor mass.
Note that B is equivalent to 10$^{51}$\,erg.\\
\textbf{b.} Same as left, but after imposing [Ca/Fe] = 0.4. The fitting procedure converges upon imposing this constraint and strongly prefers a yield model with a progenitor mass of 12\,M$_\odot$ with an explosion energy of $3 \times 10^{50}$ erg. 
}
\label{fig:yields}
\end{figure}

Given the status of PicII-503 as a second-generation star, we fit its chemical abundance pattern with yield models from first-star supernovae to constrain the properties of its progenitor. 
Specifically, we follow the procedure in ref\cite{jls+20} to perform a $\chi^2$ fit to our element abundances in Table~\ref{tab:abundances} using the supernova yield models from ref\cite{hw+10}, which span a range of progenitor masses and explosion energies (up to 10$^{52}$\,erg).
The results of this fitting procedure are shown in extended data Figure~\ref{fig:yields}, in which all models within 2$\sigma$ of the minimum $\chi^2$ value are shown.
We find that the preferred models are loosely constrained, with  progenitor masses of $<45$\,M$_\odot$, and explosion energies of  $< 2\times$10$^{51}$ \,erg.
As an exercise, we repeat the above procedure assuming [Ca/Fe] = 0.4, which is a fiducial abundance ratio for core collapse supernovae\cite{ww+95,fn+15}. 
Upon doing this, the fit strongly prefers a progenitor mass of 12\,M$_\odot$ and an explosion energy of $3\times10^{50}$\,erg.
In either case, high explosion energies and especially high progenitor masses are excluded.

\subsection*{Compilation of literature abundances}

In this section, we describe the compilation of chemical abundances shown in Figure~\ref{fig:abundances}, and compare the barium abundance of PicII-503 to CEMP stars in the literature to assess whether its carbon-enrichment could plausibly be explained by the accretion of material from an AGB star companion.
In Figure~\ref{fig:abundances}, the halo star abundances are from refs\cite{ynb+13, rpt+14, ydb+21, lam+22}, with the [Fe/H] $< -4.0$ sample supplemented by refs\cite{pfb+16, fje+19, hhc+15, aag+17, aga+17, kbf+14,pc+05, gaa+20, bcs+15, cbf+11, bcs+18, pfl+15, fcj+15, cbs+13, sab+18, aag+16, mfe+22, fwc+20, nbc+19, lmj+21, prl+21, mfc+24, plc+25}. 
The UFD star abundances are from refs\cite{kmg+08, fek+09, nwg+10, nyg+10, sfm+10, fsg+10, llb+11, gnm+13, kfa+13, fsk+14, iaa+14, rk+14, fmb+16, jfs+16, fng+16, rmb+16, hsm+17, kcs+17, cfj+18, nml+18, ssf+18, jsf+19, mhs+19, hms+20, jls+20, hms+20, wvs+23, whm+23, hsl+24}, the dwarf spheroidal galaxy abundances are refs\cite{svj+24, oyc+25, svs+24, ljs+24, sks+17, hsl+23}, and the LMC abundances are from refs\cite{cml+24, ond+24}.
When citing the number of stars in UFDs with metallicity measurements ($> 450$), we update the number from ref\cite{s+19} with recent work from refs\cite{blp+23,csl+23,ccg+25, cfs+21, csf+22, fcb+19, hsl+24, hlp+24, jlp+21, sle+20, sjr+23, ccd+25} that is listed in the Local Volume Database\cite{pace+24}.

One pathway for carbon-enhancement to occur in a star is through the accretion of material from an AGB companion; this process also leaves an enhancement in material formed from the slow neutron capture process (s-process).
These stars are known as CEMP-s stars, and typically show [Ba/Fe] $> 1.0$.
Since neither Ba nor Fe is detected in PicII-503, we perform additional tests to gauge whether this object falls in the category of a CEMP-s star.
We first select all stars with detected barium, strontium, iron, and carbon in the JINAbase compilation of metal-poor stars\cite{af+18}.
We find that all CEMP-s in this compilation (following [C/Fe] $> 1.0$, [Ba/Fe] $> 1.0$; the criteria for CEMP-s classification\cite{ybp+16}) have [C/Ba] $< 2.0$, with the vast majority having [C/Ba] $< 1.0$. 
Our abundances for PicII-503 show [C/Ba] $> 2.4$, implying the carbon enhancement does not have a consistent level of barium enhancement for the star to be a CEMP-s star. 
As further corroboration of this, we note that other CEMP-s stars in the literature tend to have high absolute carbon abundances of A(C) = 7.96$\pm0.43$ and the lowest metallicity normal CEMP stars have A(C)$\sim6.8$ (Groups I and III, respectively in ref\cite{ybp+16}). 
The absolute carbon abundance of PicII-503 is A(C) = 6.99$\pm0.38$, preferring a CEMP status that is not CEMP-s, and therefore a carbon enhancement that does not originate from an AGB companion.

\subsection*{Data availability} 
The spectra of PicII-503 are available from the corresponding author upon request. 
All X-Shooter spectra will be publicly accessible from the ESO archive (http://archive.eso.org) on March 30, 2026.

\subsection*{Code availability} The stellar synthesis code MOOG and the analysis package SMHR that were used to analyze this data can be retrieved from https://github.com/alexji/moog17scat and https://github.com/andycasey/smhr. 
The velocity analysis of the MagE and X-Shooter spectra is from the authors' implementation of code described in the methods that is straightforward to replicate, and is available from the corresponding author upon request. 
In addition to references in the publication, this work used the following packages: Astropy \cite{astropy, astropy2}, NumPy \cite{numpy}, SciPy \cite{scipy}, Matplotlib \citep{Hunter+07}, emcee \citep{emcee1, emcee2}.

\textbf{Acknowledgements} 
A.C. is supported by a Brinson Prize Fellowship.
The work of V.M.P. and Y.C. is supported by NOIRLab, which is managed by the Association of Universities for Research in Astronomy (AURA) under a cooperative agreement with the U.S. National Science Foundation. W.C. gratefully acknowledges support from a Gruber Science Fellowship at Yale University.  This material is based upon work supported by the National Science Foundation Graduate
Research Fellowship Program under Grant No. DGE2139841.

This project used data obtained with the Dark Energy Camera (DECam), which was constructed by the Dark Energy Survey (DES) collaboration. 
Funding for the DES Projects has been provided by the US Department of Energy, the U.S. National Science Foundation, the Ministry of Science and Education of Spain, the Science and Technology Facilities Council of the United Kingdom, the Higher Education Funding Council for England, the National Center for Supercomputing Applications at the University of Illinois at Urbana–Champaign, the Kavli Institute for Cosmological Physics at the University of Chicago, the Center for Cosmology and Astro-Particle Physics at the Ohio State University, the Mitchell Institute for Fundamental Physics and Astronomy at Texas A\&M University, Financiadora de Estudos e Projetos, Fundação Carlos Chagas Filho de Amparo à Pesquisa do Estado do Rio de Janeiro, Conselho 12 Nacional de Desenvolvimento Científico e Tecnológico and the Ministério da Ciência, Tecnologia e Inovação, the Deutsche Forschungsgemeinschaft and the Collaborating Institutions in the Dark Energy Survey.

The Collaborating Institutions are Argonne National Laboratory, the University of California at Santa Cruz, the University of Cambridge, Centro de Investigaciones Enérgeticas, Medioambientales y Tecnológicas–Madrid, the University of Chicago, University College London, the DES-Brazil Consortium, the University of Edinburgh, the Eidgenössische Technische Hochschule (ETH) Zürich, Fermi National Accelerator Laboratory, the University of Illinois at Urbana-Champaign, the Institut de Ciències de l’Espai (IEEC/CSIC), the Institut de Física d’Altes Energies, Lawrence Berkeley National Laboratory, the Ludwig-Maximilians Universität München and the associated Excellence Cluster Universe, the University of Michigan, NSF NOIRLab, the University of Nottingham, the Ohio State University, the OzDES Membership Consortium, the University of Pennsylvania, the University of Portsmouth, SLAC National Accelerator Laboratory, Stanford University, the University of Sussex, and Texas A\&M University.

The results are based on observations at NSF Cerro Tololo Inter-American Observatory, NSF NOIRLab (NOIRLab Prop. ID 2019A-0305; PI: Alex Drlica-Wagner, and NOIRLab Prop. ID 2023B-646244; PI: Anirudh Chiti), which is managed by the Association of Universities for Research in Astronomy (AURA) under a cooperative agreement with the U.S. National Science Foundation.

Our spectroscopic data were gathered using the 6.5-m Magellan Baade telescope located at Las Campanas Observatory, Chile and the ESO Very Large Telescope.
Specifically, based on observations collected at the European Southern Observatory under ESO programme 114.28L2.001. This work made use of NASAs Astrophysics Data System Bibliographic Services, the SIMBAD database\cite{woe+00}, operated at CDS, Strasbourg, France. 

The DELVE project is partially supported by the NASA Fermi Guest Investigator Program Cycle 9 No. 91201.
This work is partially supported by Fermilab LDRD project L2019-011. 
This material is based upon work supported by the National Science Foundation under Grant No. AST-2108168, AST-2108169, AST-2307126, and AST-2407526.

Fermilab is managed by FermiForward Discovery Group, LLC under Contract No. 89243024CSC000002 with the U.S. Department of Energy, Office of Science, Office of High Energy Physics. The United States Government retains and the publisher, by accepting the article for publication, acknowledges that the United States Government retains a non-exclusive, paid-up, irrevocable, world-wide license to publish or reproduce the published form of this manuscript, or allow others to do so, for United States Government purposes.

This work has made use of data from the European Space Agency (ESA) mission Gaia (https://www. cosmos.esa.int/gaia), processed by the Gaia Data Processing and Analysis Consortium (DPAC, https://www. cosmos.esa.int/web/gaia/dpac/consortium). Funding for the DPAC has been provided by national institutions, in particular the institutions participating in the Gaia Multilateral Agreement.

\noindent\textbf{Author contributions} A.C. led the photometering and subsequent processing of the MAGIC data to catalog generation, photometric metallicity derivation, spectroscopic targeting, MagE observations, derivation of chemical abundances, radial velocity analyses, and drafted the interpretation and text. V.M.P. assisted with the chemical abundance analysis, initial interpretation, and text; A.B.P. performed the membership analysis, and assisted with the initial interpretation, and text; A.P.J. assisted with the chemical abundance analysis, helped scope the interpretation, and assisted with the text; D.S.P. assisted with the MagE observations, the initial interpretation, and text; W.C., G.L., and G.S.S. assisted with the initial interpretation, and text; A.D.W. served as the internal referee within the collaboration; A.R.W led the acquisition of the Ca II HK filter on DECam; all authors provided feedback on the text, interpretation, and/or significantly contributed to the observations and infrastructure of the MAGIC and DELVE surveys. 

\noindent\textbf{Competing interests} The authors declare no competing financial or non-financial interests.

\noindent\textbf{Materials \& Correspondence}
Correspondence \& request for materials should be sent to Anirudh Chiti (achiti@uchicago.edu). 

\bibliography{sn-bibliography}

\end{document}